\documentstyle[12pt]{article}
\newcommand{\be}{\begin{equation}}
\newcommand{\ee}{\end{equation}}
\newcommand{\bea}{\begin{eqnarray}}
\newcommand{\eea}{\end{eqnarray}}

\begin{document}

\immediate\write16{<WARNING: FEYNMAN macros work only with emTeX-dvivers
                    (dviscr.exe, dvihplj.exe, dvidot.exe, etc.) >}
\newdimen\Lengthunit
\newcount\Nhalfperiods
\Lengthunit = 1.5cm
\Nhalfperiods = 9
\catcode`\*=11
\newdimen\L*   \newdimen\d*   \newdimen\d**
\newdimen\dm*  \newdimen\dd*  \newdimen\dt*
\newdimen\a*   \newdimen\b*   \newdimen\c*
\newdimen\a**  \newdimen\b**
\newdimen\xL*  \newdimen\yL*
\newcount\k*   \newcount\l*   \newcount\m*
\newcount\n*   \newcount\dn*  \newcount\r*
\newcount\N*   \newcount\*one \newcount\*two  \*one=1 \*two=2
\newcount\*ths \*ths=1000
\def\GRAPH(hsize=#1)#2{\hbox to #1\Lengthunit{#2\hss}}
\def\Linewidth#1{\special{em:linewidth #1}}
\Linewidth{.4pt}
\def\sm*{\special{em:moveto}}
\def\sl*{\special{em:lineto}}
\newbox\spm*   \newbox\spl*
\setbox\spm*\hbox{\sm*}
\setbox\spl*\hbox{\sl*}
\def\mov#1(#2,#3)#4{\rlap{\L*=#1\Lengthunit\kern#2\L*\raise#3\L*\hbox{#4}}}
\def\smov#1(#2,#3)#4{\rlap{\L*=#1\Lengthunit
\xL*=\xscale\L*\yL*=\yscale\L*\kern#2\xL*\raise#3\yL*\hbox{#4}}}
\def\mov*(#1,#2)#3{\rlap{\kern#1\raise#2\hbox{#3}}}
\def\lin#1(#2,#3){\rlap{\sm*\mov#1(#2,#3){\sl*}}}
\def\arr*(#1,#2,#3){\mov*(#1\dd*,#1\dt*){%
\sm*\mov*(#2\dd*,#2\dt*){\mov*(#3\dt*,-#3\dd*){\sl*}}%
\sm*\mov*(#2\dd*,#2\dt*){\mov*(-#3\dt*,#3\dd*){\sl*}}}}
\def\arrow#1(#2,#3){\rlap{\lin#1(#2,#3)\mov#1(#2,#3){%
\d**=-.012\Lengthunit\dd*=#2\d**\dt*=#3\d**%
\arr*(1,10,4)\arr*(3,8,4)\arr*(4.8,4.2,3)}}}
\def\arrlin#1(#2,#3){\rlap{\L*=#1\Lengthunit\L*=.5\L*%
\lin#1(#2,#3)\mov*(#2\L*,#3\L*){\arrow.1(#2,#3)}}}
\def\dasharrow#1(#2,#3){\rlap{%
{\Lengthunit=0.9\Lengthunit\dashlin#1(#2,#3)\mov#1(#2,#3){\sm*}}%
\mov#1(#2,#3){\sl*\d**=-.012\Lengthunit\dd*=#2\d**\dt*=#3\d**%
\arr*(1,10,4)\arr*(3,8,4)\arr*(4.8,4.2,3)}}}
\def\clap#1{\hbox to 0pt{\hss #1\hss}}
\def\ind(#1,#2)#3{\rlap{%
\d*=.1\Lengthunit\kern#1\d*\raise#2\d*\hbox{\lower2pt\clap{$#3$}}}}
\def\sh*(#1,#2)#3{\rlap{%
\dm*=\the\n*\d**\xL*=\xscale\dm*\yL*=\yscale\dm*
\kern#1\xL*\raise#2\yL*\hbox{#3}}}
\def\calcnum*#1(#2,#3){\a*=1000sp\b*=1000sp\a*=#2\a*\b*=#3\b*%
\ifdim\a*<0pt\a*-\a*\fi\ifdim\b*<0pt\b*-\b*\fi%
\ifdim\a*>\b*\c*=.96\a*\advance\c*.4\b*%
\else\c*=.96\b*\advance\c*.4\a*\fi%
\k*\a*\multiply\k*\k*\l*\b*\multiply\l*\l*%
\m*\k*\advance\m*\l*\n*\c*\r*\n*\multiply\n*\n*%
\dn*\m*\advance\dn*-\n*\divide\dn*2\divide\dn*\r*%
\advance\r*\dn*%
\c*=\the\Nhalfperiods5sp\c*=#1\c*\ifdim\c*<0pt\c*-\c*\fi%
\multiply\c*\r*\N*\c*\divide\N*10000}
\def\dashlin#1(#2,#3){\rlap{\calcnum*#1(#2,#3)%
\d**=#1\Lengthunit\ifdim\d**<0pt\d**-\d**\fi%
\divide\N*2\multiply\N*2\advance\N*1%
\divide\d**\N*\sm*\n*\*one\sh*(#2,#3){\sl*}%
\loop\advance\n*\*one\sh*(#2,#3){\sm*}\advance\n*\*one\sh*(#2,#3){\sl*}%
\ifnum\n*<\N*\repeat}}
\def\dashdotlin#1(#2,#3){\rlap{\calcnum*#1(#2,#3)%
\d**=#1\Lengthunit\ifdim\d**<0pt\d**-\d**\fi%
\divide\N*2\multiply\N*2\advance\N*1\multiply\N*2%
\divide\d**\N*\sm*\n*\*two\sh*(#2,#3){\sl*}\loop%
\advance\n*\*one\sh*(#2,#3){\kern-1.48pt\lower.5pt\hbox{\rm.}}%
\advance\n*\*one\sh*(#2,#3){\sm*}%
\advance\n*\*two\sh*(#2,#3){\sl*}\ifnum\n*<\N*\repeat}}
\def\shl*(#1,#2)#3{\kern#1#3\lower#2#3\hbox{\unhcopy\spl*}}
\def\trianglin#1(#2,#3){\rlap{\toks0={#2}\toks1={#3}\calcnum*#1(#2,#3)%
\dd*=.57\Lengthunit\dd*=#1\dd*\divide\dd*\N*%
\d**=#1\Lengthunit\ifdim\d**<0pt\d**-\d**\fi%
\multiply\N*2\divide\d**\N*\advance\N*-1\sm*\n*\*one\loop%
\shl**{\dd*}\dd*-\dd*\advance\n*2%
\ifnum\n*<\N*\repeat\n*\N*\advance\n*1\shl**{0pt}}}
\def\wavelin#1(#2,#3){\rlap{\toks0={#2}\toks1={#3}\calcnum*#1(#2,#3)%
\dd*=.23\Lengthunit\dd*=#1\dd*\divide\dd*\N*%
\d**=#1\Lengthunit\ifdim\d**<0pt\d**-\d**\fi%
\multiply\N*4\divide\d**\N*\sm*\n*\*one\loop%
\shl**{\dd*}\dt*=1.3\dd*\advance\n*1%
\shl**{\dt*}\advance\n*\*one%
\shl**{\dd*}\advance\n*\*two%
\dd*-\dd*\ifnum\n*<\N*\repeat\n*\N*\shl**{0pt}}}
\def\w*lin(#1,#2){\rlap{\toks0={#1}\toks1={#2}\d**=\Lengthunit\dd*=-.12\d**%
\N*8\divide\d**\N*\sm*\n*\*one\loop%
\shl**{\dd*}\dt*=1.3\dd*\advance\n*\*one%
\shl**{\dt*}\advance\n*\*one%
\shl**{\dd*}\advance\n*\*one%
\shl**{0pt}\dd*-\dd*\advance\n*1\ifnum\n*<\N*\repeat}}
\def\l*arc(#1,#2)[#3][#4]{\rlap{\toks0={#1}\toks1={#2}\d**=\Lengthunit%
\dd*=#3.037\d**\dd*=#4\dd*\dt*=#3.049\d**\dt*=#4\dt*\ifdim\d**>16mm%
\d**=.25\d**\n*\*one\shl**{-\dd*}\n*\*two\shl**{-\dt*}\n*3\relax%
\shl**{-\dd*}\n*4\relax\shl**{0pt}\else\ifdim\d**>5mm%
\d**=.5\d**\n*\*one\shl**{-\dt*}\n*\*two\shl**{0pt}%
\else\n*\*one\shl**{0pt}\fi\fi}}
\def\d*arc(#1,#2)[#3][#4]{\rlap{\toks0={#1}\toks1={#2}\d**=\Lengthunit%
\dd*=#3.037\d**\dd*=#4\dd*\d**=.25\d**\sm*\n*\*one\shl**{-\dd*}%
\n*3\relax\sh*(#1,#2){\xL*=\xscale\dd*\yL*=\yscale\dd*
\kern#2\xL*\lower#1\yL*\hbox{\sm*}}%
\n*4\relax\shl**{0pt}}}
\def\arc#1[#2][#3]{\rlap{\Lengthunit=#1\Lengthunit%
\sm*\l*arc(#2.1914,#3.0381)[#2][#3]%
\smov(#2.1914,#3.0381){\l*arc(#2.1622,#3.1084)[#2][#3]}%
\smov(#2.3536,#3.1465){\l*arc(#2.1084,#3.1622)[#2][#3]}%
\smov(#2.4619,#3.3086){\l*arc(#2.0381,#3.1914)[#2][#3]}}}
\def\dasharc#1[#2][#3]{\rlap{\Lengthunit=#1\Lengthunit%
\d*arc(#2.1914,#3.0381)[#2][#3]%
\smov(#2.1914,#3.0381){\d*arc(#2.1622,#3.1084)[#2][#3]}%
\smov(#2.3536,#3.1465){\d*arc(#2.1084,#3.1622)[#2][#3]}%
\smov(#2.4619,#3.3086){\d*arc(#2.0381,#3.1914)[#2][#3]}}}
\def\wavearc#1[#2][#3]{\rlap{\Lengthunit=#1\Lengthunit%
\w*lin(#2.1914,#3.0381)%
\smov(#2.1914,#3.0381){\w*lin(#2.1622,#3.1084)}%
\smov(#2.3536,#3.1465){\w*lin(#2.1084,#3.1622)}%
\smov(#2.4619,#3.3086){\w*lin(#2.0381,#3.1914)}}}
\def\shl**#1{\c*=\the\n*\d**\d*=#1%
\a*=\the\toks0\c*\b*=\the\toks1\d*\advance\a*-\b*%
\b*=\the\toks1\c*\d*=\the\toks0\d*\advance\b*\d*%
\a*=\xscale\a*\b*=\yscale\b*%
\raise\b*\rlap{\kern\a*\unhcopy\spl*}}
\def\wlin*#1(#2,#3)[#4]{\rlap{\toks0={#2}\toks1={#3}%
\c*=#1\l*\c*\c*=.01\Lengthunit\m*\c*\divide\l*\m*%
\c*=\the\Nhalfperiods5sp\multiply\c*\l*\N*\c*\divide\N*\*ths%
\divide\N*2\multiply\N*2\advance\N*1%
\dd*=.002\Lengthunit\dd*=#4\dd*\multiply\dd*\l*\divide\dd*\N*%
\d**=#1\multiply\N*4\divide\d**\N*\sm*\n*\*one\loop%
\shl**{\dd*}\dt*=1.3\dd*\advance\n*\*one%
\shl**{\dt*}\advance\n*\*one%
\shl**{\dd*}\advance\n*\*two%
\dd*-\dd*\ifnum\n*<\N*\repeat\n*\N*\shl**{0pt}}}
\def\wavebox#1{\setbox0\hbox{#1}%
\a*=\wd0\advance\a*14pt\b*=\ht0\advance\b*\dp0\advance\b*14pt%
\hbox{\kern9pt%
\mov*(0pt,\ht0){\mov*(-7pt,7pt){\wlin*\a*(1,0)[+]\wlin*\b*(0,-1)[-]}}%
\mov*(\wd0,-\dp0){\mov*(7pt,-7pt){\wlin*\a*(-1,0)[+]\wlin*\b*(0,1)[-]}}%
\box0\kern9pt}}
\def\rectangle#1(#2,#3){%
\lin#1(#2,0)\lin#1(0,#3)\mov#1(0,#3){\lin#1(#2,0)}\mov#1(#2,0){\lin#1(0,#3)}}
\def\dashrectangle#1(#2,#3){\dashlin#1(#2,0)\dashlin#1(0,#3)%
\mov#1(0,#3){\dashlin#1(#2,0)}\mov#1(#2,0){\dashlin#1(0,#3)}}
\def\waverectangle#1(#2,#3){\L*=#1\Lengthunit\a*=#2\L*\b*=#3\L*%
\ifdim\a*<0pt\a*-\a*\def\x*{-1}\else\def\x*{1}\fi%
\ifdim\b*<0pt\b*-\b*\def\y*{-1}\else\def\y*{1}\fi%
\wlin*\a*(\x*,0)[-]\wlin*\b*(0,\y*)[+]%
\mov#1(0,#3){\wlin*\a*(\x*,0)[+]}\mov#1(#2,0){\wlin*\b*(0,\y*)[-]}}
\def\calcparab*{%
\ifnum\n*>\m*\k*\N*\advance\k*-\n*\else\k*\n*\fi%
\a*=\the\k* sp\a*=10\a*\b*\dm*\advance\b*-\a*\k*\b*%
\a*=\the\*ths\b*\divide\a*\l*\multiply\a*\k*%
\divide\a*\l*\k*\*ths\r*\a*\advance\k*-\r*%
\dt*=\the\k*\L*}
\def\arcto#1(#2,#3)[#4]{\rlap{\toks0={#2}\toks1={#3}\calcnum*#1(#2,#3)%
\dm*=135sp\dm*=#1\dm*\d**=#1\Lengthunit\ifdim\dm*<0pt\dm*-\dm*\fi%
\multiply\dm*\r*\a*=.3\dm*\a*=#4\a*\ifdim\a*<0pt\a*-\a*\fi%
\advance\dm*\a*\N*\dm*\divide\N*10000%
\divide\N*2\multiply\N*2\advance\N*1%
\L*=-.25\d**\L*=#4\L*\divide\d**\N*\divide\L*\*ths%
\m*\N*\divide\m*2\dm*=\the\m*5sp\l*\dm*%
\sm*\n*\*one\loop\calcparab*\shl**{-\dt*}%
\advance\n*1\ifnum\n*<\N*\repeat}}
\def\arrarcto#1(#2,#3)[#4]{\L*=#1\Lengthunit\L*=.54\L*%
\arcto#1(#2,#3)[#4]\mov*(#2\L*,#3\L*){\d*=.457\L*\d*=#4\d*\d**-\d*%
\mov*(#3\d**,#2\d*){\arrow.02(#2,#3)}}}
\def\dasharcto#1(#2,#3)[#4]{\rlap{\toks0={#2}\toks1={#3}\calcnum*#1(#2,#3)%
\dm*=\the\N*5sp\a*=.3\dm*\a*=#4\a*\ifdim\a*<0pt\a*-\a*\fi%
\advance\dm*\a*\N*\dm*%
\divide\N*20\multiply\N*2\advance\N*1\d**=#1\Lengthunit%
\L*=-.25\d**\L*=#4\L*\divide\d**\N*\divide\L*\*ths%
\m*\N*\divide\m*2\dm*=\the\m*5sp\l*\dm*%
\sm*\n*\*one\loop%
\calcparab*\shl**{-\dt*}\advance\n*1%
\ifnum\n*>\N*\else\calcparab*%
\sh*(#2,#3){\kern#3\dt*\lower#2\dt*\hbox{\sm*}}\fi%
\advance\n*1\ifnum\n*<\N*\repeat}}
\def\*shl*#1{%
\c*=\the\n*\d**\advance\c*#1\a**\d*\dt*\advance\d*#1\b**%
\a*=\the\toks0\c*\b*=\the\toks1\d*\advance\a*-\b*%
\b*=\the\toks1\c*\d*=\the\toks0\d*\advance\b*\d*%
\raise\b*\rlap{\kern\a*\unhcopy\spl*}}
\def\calcnormal*#1{%
\b**=10000sp\a**\b**\k*\n*\advance\k*-\m*%
\multiply\a**\k*\divide\a**\m*\a**=#1\a**\ifdim\a**<0pt\a**-\a**\fi%
\ifdim\a**>\b**\d*=.96\a**\advance\d*.4\b**%
\else\d*=.96\b**\advance\d*.4\a**\fi%
\d*=.01\d*\r*\d*\divide\a**\r*\divide\b**\r*%
\ifnum\k*<0\a**-\a**\fi\d*=#1\d*\ifdim\d*<0pt\b**-\b**\fi%
\k*\a**\a**=\the\k*\dd*\k*\b**\b**=\the\k*\dd*}
\def\wavearcto#1(#2,#3)[#4]{\rlap{\toks0={#2}\toks1={#3}\calcnum*#1(#2,#3)%
\c*=\the\N*5sp\a*=.4\c*\a*=#4\a*\ifdim\a*<0pt\a*-\a*\fi%
\advance\c*\a*\N*\c*\divide\N*20\multiply\N*2\advance\N*-1\multiply\N*4%
\d**=#1\Lengthunit\dd*=.012\d**\ifdim\d**<0pt\d**-\d**\fi\L*=.25\d**%
\divide\d**\N*\divide\dd*\N*\L*=#4\L*\divide\L*\*ths%
\m*\N*\divide\m*2\dm*=\the\m*0sp\l*\dm*%
\sm*\n*\*one\loop\calcnormal*{#4}\calcparab*%
\*shl*{1}\advance\n*\*one\calcparab*%
\*shl*{1.3}\advance\n*\*one\calcparab*%
\*shl*{1}\advance\n*2%
\dd*-\dd*\ifnum\n*<\N*\repeat\n*\N*\shl**{0pt}}}
\def\triangarcto#1(#2,#3)[#4]{\rlap{\toks0={#2}\toks1={#3}\calcnum*#1(#2,#3)%
\c*=\the\N*5sp\a*=.4\c*\a*=#4\a*\ifdim\a*<0pt\a*-\a*\fi%
\advance\c*\a*\N*\c*\divide\N*20\multiply\N*2\advance\N*-1\multiply\N*2%
\d**=#1\Lengthunit\dd*=.012\d**\ifdim\d**<0pt\d**-\d**\fi\L*=.25\d**%
\divide\d**\N*\divide\dd*\N*\L*=#4\L*\divide\L*\*ths%
\m*\N*\divide\m*2\dm*=\the\m*0sp\l*\dm*%
\sm*\n*\*one\loop\calcnormal*{#4}\calcparab*%
\*shl*{1}\advance\n*2%
\dd*-\dd*\ifnum\n*<\N*\repeat\n*\N*\shl**{0pt}}}
\def\hr*#1{\clap{\xL*=\xscale\Lengthunit\vrule width#1\xL* height.1pt}}
\def\shade#1[#2]{\rlap{\Lengthunit=#1\Lengthunit%
\smov(0,#2.05){\hr*{.994}}\smov(0,#2.1){\hr*{.980}}%
\smov(0,#2.15){\hr*{.953}}\smov(0,#2.2){\hr*{.916}}%
\smov(0,#2.25){\hr*{.867}}\smov(0,#2.3){\hr*{.798}}%
\smov(0,#2.35){\hr*{.715}}\smov(0,#2.4){\hr*{.603}}%
\smov(0,#2.45){\hr*{.435}}}}
\def\dshade#1[#2]{\rlap{%
\Lengthunit=#1\Lengthunit\if#2-\def\t*{+}\else\def\t*{-}\fi%
\smov(0,\t*.025){%
\smov(0,#2.05){\hr*{.995}}\smov(0,#2.1){\hr*{.988}}%
\smov(0,#2.15){\hr*{.969}}\smov(0,#2.2){\hr*{.937}}%
\smov(0,#2.25){\hr*{.893}}\smov(0,#2.3){\hr*{.836}}%
\smov(0,#2.35){\hr*{.760}}\smov(0,#2.4){\hr*{.662}}%
\smov(0,#2.45){\hr*{.531}}\smov(0,#2.5){\hr*{.320}}}}}
\def\vdot{\rlap{\kern-1.9pt\lower1.8pt\hbox{$\scriptstyle\bullet$}}}
\def\vtimes{\rlap{\kern-3pt\lower1.8pt\hbox{$\scriptstyle\times$}}}
\def\vDot{\rlap{\kern-2.3pt\lower2.7pt\hbox{$\bullet$}}}
\def\vTimes{\rlap{\kern-3.6pt\lower2.4pt\hbox{$\times$}}}
\catcode`\*=12
\newcount\CatcodeOfAtSign
\CatcodeOfAtSign=\the\catcode`\@
\catcode`\@=11
\newcount\n@ast
\def\n@ast@#1{\n@ast0\relax\get@ast@#1\end}
\def\get@ast@#1{\ifx#1\end\let\next\relax\else%
\ifx#1*\advance\n@ast1\fi\let\next\get@ast@\fi\next}
\newif\if@up \newif\if@dwn
\def\up@down@#1{\@upfalse\@dwnfalse%
\if#1u\@uptrue\fi\if#1U\@uptrue\fi\if#1+\@uptrue\fi%
\if#1d\@dwntrue\fi\if#1D\@dwntrue\fi\if#1-\@dwntrue\fi}
\def\halfcirc#1(#2)[#3]{{\Lengthunit=#2\Lengthunit\up@down@{#3}%
\if@up\smov(0,.5){\arc[-][-]\arc[+][-]}\fi%
\if@dwn\smov(0,-.5){\arc[-][+]\arc[+][+]}\fi%
\def\lft{\smov(0,.5){\arc[-][-]}\smov(0,-.5){\arc[-][+]}}%
\def\rght{\smov(0,.5){\arc[+][-]}\smov(0,-.5){\arc[+][+]}}%
\if#3l\lft\fi\if#3L\lft\fi\if#3r\rght\fi\if#3R\rght\fi%
\n@ast@{#1}%
\ifnum\n@ast>0\if@up\shade[+]\fi\if@dwn\shade[-]\fi\fi%
\ifnum\n@ast>1\if@up\dshade[+]\fi\if@dwn\dshade[-]\fi\fi}}
\def\halfdashcirc(#1)[#2]{{\Lengthunit=#1\Lengthunit\up@down@{#2}%
\if@up\smov(0,.5){\dasharc[-][-]\dasharc[+][-]}\fi%
\if@dwn\smov(0,-.5){\dasharc[-][+]\dasharc[+][+]}\fi%
\def\lft{\smov(0,.5){\dasharc[-][-]}\smov(0,-.5){\dasharc[-][+]}}%
\def\rght{\smov(0,.5){\dasharc[+][-]}\smov(0,-.5){\dasharc[+][+]}}%
\if#2l\lft\fi\if#2L\lft\fi\if#2r\rght\fi\if#2R\rght\fi}}
\def\halfwavecirc(#1)[#2]{{\Lengthunit=#1\Lengthunit\up@down@{#2}%
\if@up\smov(0,.5){\wavearc[-][-]\wavearc[+][-]}\fi%
\if@dwn\smov(0,-.5){\wavearc[-][+]\wavearc[+][+]}\fi%
\def\lft{\smov(0,.5){\wavearc[-][-]}\smov(0,-.5){\wavearc[-][+]}}%
\def\rght{\smov(0,.5){\wavearc[+][-]}\smov(0,-.5){\wavearc[+][+]}}%
\if#2l\lft\fi\if#2L\lft\fi\if#2r\rght\fi\if#2R\rght\fi}}
\def\Circle#1(#2){\halfcirc#1(#2)[u]\halfcirc#1(#2)[d]\n@ast@{#1}%
\ifnum\n@ast>0\clap{%
\dimen0=\xscale\Lengthunit\vrule width#2\dimen0 height.1pt}\fi}
\def\wavecirc(#1){\halfwavecirc(#1)[u]\halfwavecirc(#1)[d]}
\def\dashcirc(#1){\halfdashcirc(#1)[u]\halfdashcirc(#1)[d]}
%
\def\xscale{1}
\def\yscale{1}
\def\Ellipse#1(#2)[#3,#4]{\def\xscale{#3}\def\yscale{#4}%
\Circle#1(#2)\def\xscale{1}\def\yscale{1}}
\def\dashEllipse(#1)[#2,#3]{\def\xscale{#2}\def\yscale{#3}%
\dashcirc(#1)\def\xscale{1}\def\yscale{1}}
\def\waveEllipse(#1)[#2,#3]{\def\xscale{#2}\def\yscale{#3}%
\wavecirc(#1)\def\xscale{1}\def\yscale{1}}
\def\halfEllipse#1(#2)[#3][#4,#5]{\def\xscale{#4}\def\yscale{#5}%
\halfcirc#1(#2)[#3]\def\xscale{1}\def\yscale{1}}
\def\halfdashEllipse(#1)[#2][#3,#4]{\def\xscale{#3}\def\yscale{#4}%
\halfdashcirc(#1)[#2]\def\xscale{1}\def\yscale{1}}
\def\halfwaveEllipse(#1)[#2][#3,#4]{\def\xscale{#3}\def\yscale{#4}%
\halfwavecirc(#1)[#2]\def\xscale{1}\def\yscale{1}}
\catcode`\@=\the\CatcodeOfAtSign

\begin{titlepage}
\thispagestyle{empty}

\begin{flushleft}
IASSNS-HEP-97/6\\
UPR-733T\\
JINR E2-97-82\\
ITP-UH-09/97\\
hep-th/9703147
\end{flushleft}

\begin{flushright}
{\it To the memory of}
{\it Professor V.I. Ogievetsky}
\end{flushright}
\vspace{5mm}

\begin{center}
{\large\bf Effective Action of the N = 2 Maxwell
Multiplet in Harmonic Superspace}
\end{center}
\vspace{5mm}

\begin{center}
I. L. Buchbinder$\,^{a)}$, E. I. Buchbinder$\,^{b)}$, E. A. Ivanov$\,^{b)}$,
S. M. Kuzenko$\,$\footnote{Alexander von Humboldt Research Fellow.
On leave from Department of Quantum Field Theory, Tomsk State University,
Tomsk 634050, Russia}${}^{,c)}$, B. A. Ovrut$\,^{d)}$
\end{center}

\begin{itemize}
\item[${}^{a)}$] \footnotesize{{\it Department of Theoretical Physics,
Tomsk State Pedagogical University, Tomsk 634041, Russia}}

\item[${}^{b)}$] \footnotesize{{\it Bogoliubov Laboratory of Theoretical
Physics, Joint Institute for Nuclear Research, \\
Dubna, Moscow Region, 141980, Russia}}

\item[${}^{c)}$] \footnotesize{{\it Institut f\"ur Theoretische Physik,
Universit\"at Hannover,
Appelstr. 2, 30167 Hannover, Germany}}

\item[${}^{d)}$] \footnotesize{{\it School of Natural Sciences,
Institute for Advanced Study, Olden Lane, Princeton, NJ 08540, USA\\
Department of Physics, University of
Pennsylvania, Philadelphia, PA 19104-6396, USA}}
\end{itemize}

\begin{abstract}
We present, in the $N=2$, $D=4$ harmonic superspace formalism,
a general method for constructing the off-shell effective action
of an $N=2$ abelian gauge superfield coupled to
matter hypermultiplets.
Using manifestly $N=2$  supersymmetric harmonic supergraph techniques,
we calculate the low-energy corrections to the renormalized one-loop
effective action in terms
of $N=2$ (anti)chiral superfield strengths.
For a harmonic gauge prepotential with vanishing vacuum expectation value,
corresponding to massless hypermultiplets, the only non-trivial radiative
corrections to appear are non-holomorphic. For a prepotential with 
non-zero vacuum value, which
breaks the $U(1)$-factor in the $N=2$ supersymmetry automorphism group and
corresponds to massive hypermultiplets, only non-trivial holomorphic
corrections arise at leading order. 
These holomorphic contribution are consistent with
Seiberg's quantum correction to the effective action, while the
first non-holomorphic contribution in the massless case is the $N=2$ 
supersymmetrization of the
Heisenberg-Euler effective Lagrangian.
\end{abstract}
\vfill

\end{titlepage}

\newpage
\setcounter{page}{1}

\noindent
$N=2$ supersymmetric field theories possess remarkable
properties both at the classical and quantum levels. Applications
of $N=2$ supersymmetry range from
superstring theory to topological field theory, supergauge models
and special geometry (see \cite{fs} for a modern review). Although
the theory of $N=2$ supersymmetry has a long history, it still has properties yet
to be explored.

During the last few years, quantum aspects of $N=2$
supersymmetric theories have excited considerable interest. This interest
was inspired by the seminal papers of Seiberg and Witten \cite{sw} where
the non-perturbative contribution to the low-energy effective action of
the $N=2$, $SU(2)$ super Yang-Mills model were calculated exactly.
The content of Refs. \cite{sw} is essentially based on the structure
of the low-energy effective action proposed in Ref. \cite{s} (see also \cite{g}).

A key element of the whole approach of \cite{sw} is the statement that
the leading contribution to the low-energy effective action of $N=2$
super Yang-Mills theory is represented by a single holomorphic function
of the $N=2$ chiral superfield strength $W$. A detailed investigation of this
statement, and the calculation of non-leading contributions to the
low-energy effective action, have been undertaken in recent papers
[5--9]\footnote{As it was noted in \cite{h}, such non-leading
contributions are described in terms of a real function of $W$
and its conjugate ${\bar W}$.}.

As is well known, an adequate description of quantum $N$-extended
supersymmetric field theories can be achieved in terms of
unconstrained superfields given on an appropriate $N$-extended superspace.
However, the analysis of Refs. [5--9], as well as the
main statement of Ref. \cite{s}, were based on the
formulation of $N=2$ supersymmetric theories in terms of $N=1$ superfields.
Such formulations lack manifest $N=2$ supersymmetry which, in general,
gets closed partly on-shell. Since these formulations do not 
keep $N=2$ supersymmetry manifest at all stages
of the computation, they can lead to a number of obstacles.
In this respect, the problem of
calculating the effective action of $N=2$ theories in terms of
unconstrained $N=2$ superfields appears to be of importance.

$N=2$ supersymmetric theories can be
formulated in standard $N=2$ superspace in terms of constrained
superfields. For a special $N=2$ matter multiplet (the so called relaxed
hypermultiplet \cite{hst}) and the gauge multiplet
the corresponding constraints were solved in \cite{hst,constr}.
However, these formulations look extremely
complicated when the interaction is switched on and, in our opinion,
are very difficult to use for the computation of the effective action.

A constructive and elegant approach to the description of theories
with extended supersymmetry is based on the concept
of harmonic superspace [13--16].
It allows one to investigate different extended supersymmetric models
naturally and simply. As to $N=2$ models, their formulation using the
harmonic superspace approach looks quite transparent.

In this letter we begin an investigation of the
quantum aspects of $N=2$, $D=4$ supersymmetric field theories
using the harmonic superspace approach. We study the low-energy structure
of the Wilsonian effective action of an abelian gauge superfield
coupled to matter superfields.

Because of $N=2$ supersymmetry and gauge invariance, which the harmonic
superspace approach allows us to keep manifest, the effective action of
the Maxwell multiplet is a non-local functional of the (anti)chiral
superfield strengths $W$ and ${\bar W}$ only. In the low-energy limit, when
only the leading contribution in the space-time
derivatives survives, we are left with a local effective superpotential
depending only on $W$ and ${\bar W}$.

\vspace{0.5cm}

The Fayet-Sohnius massless hypermultiplet is described in
harmonic superspace by an unconstrained analytic superfield
$q^+(\zeta_A,u^+,u^-)$ \cite{gikos}, where
$\zeta^M_A\equiv (x^m_A, \theta^{+\alpha},{\bar\theta^+_{\dot\alpha}})$
are the coordinates of an analytic subspace of the whole
$N=2$, $D=4$ harmonic superspace,
$\theta^+_\alpha=\theta^i_\alpha u^+_i$, ${\bar\theta^+_{\dot\alpha}}=
{\bar\theta^i_{\dot\alpha}}u^+_i$, $||u^\pm_i||\in SU(2)$, $i=1,2$.
The most characteristic feature of the superfield $q^+$ is an infinite
number of auxiliary fields coming from the harmonic expansions
in $u^+_i, u^-_i$. This is the only possible way to describe the off-shell
massless hypermultiplet within the framework of $N=2$ supersymmetry without
central charges. The $q^+$ multiplet is universal, all
known $N=2$ matter off-shell multiplets with finite numbers of auxiliary
fields (e.g., the relaxed hypermultiplet \cite{hst}) are related to it
via appropriate duality transformations \cite{gio}.

The classical action for the hypermultiplet interacting with
an abelian gauge superfield $V^{++}(\zeta_A,u^+,u^-)$ is given by

\be
S[\stackrel{\smile}{q}{}^+,q^+,V^{++}]=\int {\rm d}\zeta_A^{(-4)}
{\rm d}u\stackrel{\smile}{q}
{}^+ \nabla^{++}q^+ \;.
\label{1}
\ee
Here
${\rm d}\zeta_A^{(-4)}={\rm d}^4x_A{\rm d}^2\theta^+
{\rm d}^2{\bar\theta}^+$,
\be
\nabla^{++}=D^{++}+{\rm i}V^{++}\;,
\label{2}
\ee
and operation $\smile$ called `smile' denotes the analyticity-preserving
conjugation \cite{gikos} ($\stackrel{\smile}{q}{}^+\equiv
{\stackrel{*}{\bar q}{}^+}$).
The explicit form of the operator $D^{++}$ in the analytic basis, as well as
all relevant notation, can be found in Ref. \cite{gikos}.

The $S[\stackrel{\smile}{q}{}^+,q^+,V^{++}]$ enters as  part
of the action of $N=2$ supersymmetric electrodynamics
\be
S_{{\rm SED}}=\frac{1}{2}\int{\rm d}^4x{\rm d}^4\theta W^2+\int{\rm
d}\zeta^{(-4)}_A{\rm d}
u\stackrel{\smile}{q}{}^+(D^{++}+{\rm i}V^{++})q^+ \;.
\label{103}
\ee
The chiral gauge invariant strength $W$ and its conjugate $\bar W$ are
expressed via $V^{++}$
by the relation \cite{gikos,gios1,zup}
\be
W=-\int{\rm d}u({\bar D}^-)^2 V^{++}(x,\theta,u)\qquad
{\bar W}=-\int{\rm d}u(D^-)^2 V^{++}(x,\theta,u)
\label{104}
\ee
with $D^{\pm}_\alpha = D^i_\alpha u^\pm_i$,
${\bar D}^{\pm}_{\dot{\alpha}} = {\bar D}^i_{\dot{\alpha}} u^\pm_i$ the
spinor covariant derivatives.
For later use, we singled out in $V^{++}$ a background
part $V_0^{++}$ and write  $V^{++}= V_0^{++} + V_1^{++}$. $V_0^{++}$ possesses 
a constant strength $W_0$ and can be
chosen to be of the form
\be
-(\theta^+)^2{\bar W}_0-({\bar\theta}^+)^2W_0 \qquad W_0 = const\;.
\ee
For $V_0^{++}=0$, the hypermultiplets are massless. What happens when
$V_0^{++}=\neq 0$?
Whatever the origin of a non-vanishing $V_0^{++}$ (and $W_0$) may be, such a 
$V_0^{++}$ breaks the $U(1)$-factor in the $N=2$ superalgebra
automorphism group $U(2)$ and gives $q^+$ a mass $m=|W_0|$
via generating a central charge proportional to the
generator of gauge $U(1)$ symmetry
\footnote{The fact that the hypermultiplet becomes massive
follows from the dynamical equation $( D^{++} + {\rm i} V_0^{++})q^+ = 0$
which implies $(\Box + m^2) q^+ = 0$, where $m =|W_0|$.}.

Thus, this theory possesses
two different phases associated with two physically
different choices; $V_0^{++}=0$ and $V_0^{++}\neq 0$.
Because of the Bianchi identity
$D^{\alpha i}D_\alpha^j W={\bar D}^i_{\dot\alpha}{\bar D}^{{\dot\alpha}j}
{\bar W}$, we have
\be
\int{\rm d}^4x{\rm d}^4\theta(W_1+W_0)^2=\int{\rm d}^4x{\rm d}^4\theta
W_1^2 \;.
\ee
Thus,
$N=2$ Maxwell theory can be treated either as a theory of {\it massless}
superfields $q^+$,
$\stackrel{\smile}{q}{}^+$ coupled to gauge superfield
$V_1^{++}$ (the first phase), or as
a theory of {\it massive} $q^+$, $\stackrel{\smile}{q}{}^+$
coupled to $V^{++}_1$ (the second phase).
We will consider both phases.

Note that an abelian theory with $W_0 = const$
naturally arises as an effective theory describing the
spontaneous symmetry breaking phase in $N=2$ super
Yang-Mills theory. In this case the classical potential vanishes
at non-zero vacuum values of the scalar components of the gauge multiplet
and only a $U(1)$-factor of the gauge group survives.
In the superfield language, such a situation just corresponds to
$W_0 = const$ (see Ref. \cite{ah} for a generic discussion of
spontaneous symmetry breakdown in $N=2$ super Yang-Mills theory).

The effective action $\Gamma[V^{++}]$ of the theory (\ref{1})
is defined by the path integral
\be
{\rm e}^{{\rm i}\Gamma[V^{++}]}=\int{\cal D}
\stackrel{\smile}{q}{}^+{\cal D}q^+
\displaystyle {\rm e}^{{\rm i}S[\stackrel{\smile}{q}{}^+,q^+,V^{++}]}
\label{3}
\ee
and can be formally written as
\be
\Gamma[V^{++}]={\rm i}\,{\rm Tr}\;\ln \nabla^{++}\;.
\label{4}
\ee
We will calculate $\Gamma[V^{++}]$ starting with this relation,    
using a suitable definition of the right-hand side of (\ref{4}).

Another, basically equivalent version of the harmonic superspace
description of the massless hypermultiplet makes use of an unconstrained
analytic superfield $\omega(\zeta_A,u^+,u^-)$ \cite{gikos}.
It should be taken complex when coupled to the
Maxwell gauge superfield. 
We now show that the effective action for the $\omega$ version of the
hypermultiplet can be computed directly from the $q^{+}$ effective action
$\Gamma[V^{++}]$.
The classical
action for $\omega$ interacting with
the abelian $V^{++}$ is given by
\be
S[\stackrel{\smile}{\omega},\omega,V^{++}]=\int {\rm d}\zeta^{(-4)}_A
{\rm d}u\,\nabla^{++}\stackrel{\smile}{\omega}\nabla^{++}\omega
\label{5}
\ee
where
\be
\nabla^{++}\omega=(D^{++}+{\rm i}V^{++})\omega\ ,\qquad
\nabla^{++}\stackrel{\smile}{\omega}=(D^{++}-{\rm i}V^{++})
\stackrel{\smile}{\omega}\;.
\label{6}
\ee
The effective action $\Gamma_\omega[V^{++}]$ of the theory (\ref{5}) is
defined by
\be
{\rm e}^{{\rm i}\Gamma_\omega[V^{++}]}=\int{\cal D}\stackrel{\smile}{\omega}
{\cal D}\omega {\rm e}^{{\rm i}S[\stackrel{\smile}{\omega},\omega,V^{++}]}
\label{7}
\ee
and can formally be written as 
\be
\Gamma_\omega[V^{++}]={\rm i}\,{\rm Tr}\;\ln(\nabla^{++})^2 \;.
\label{8}
\ee
Eqs. (\ref{4}) and (\ref{8}) lead to the formal relation
\be
\Gamma_\omega[V^{++}]=2\Gamma[V^{++}]
\label{9}
\ee
which, of course, needs justification.

In order to make the above considerations more precise, we consider a theory
of two hypermultiplets $q_i$ ($i=1,2$) with the action
\be
{\tilde S}[\stackrel{\smile}{q}{}^{+i},q^+_i,V^{++}]=
\int {\rm d}\zeta^{(-4)}_A
{\rm d}u\stackrel{\smile}{q}{}^{+i}\nabla^{++}q^+_i
\label{10}
\ee
and introduce the corresponding effective action ${\tilde\Gamma[V^{++}]}$
defined by
\be
{\rm e}^{{\rm i}\tilde\Gamma[V^{++}]}=
\int{\cal D}\stackrel{\smile}{q}{}^{+i}
{\cal D}q^+_i {\rm e}^{{\rm i}{\tilde S}[
\stackrel{\smile}{q}{}^{+i},q^+_i,V^{++}]}
={\rm e}^{2{\rm i}\Gamma[V^{++}]}\;.
\label{11}
\ee
Let us also consider the following change of variables
\bea
\stackrel{\smile}{q}{}^{+i}&=&u^{+i}\stackrel{\smile}{\omega}+
u^{-i}\stackrel{\smile}{f}{}^{++}\;,\;\;
q^+_i = u^+_i\omega+u^-_i f^{++}
\label{12}
\eea
with some analytic superfields $f^{++}$, $\stackrel{\smile}{f}{}^{++}$.
Transformation (\ref{12}) has been introduced in Ref. \cite{gios2}
in order to prove the classical equivalence of the models
(\ref{10}) and (\ref{5})
at $V^{++}=0$ \footnote{To avoid
confusion, we point out that the single $q^+$ can also be traded for
a single {\it real}   $\omega$ hypermultiplet via eq. (\ref{12}) with
$\stackrel{\smile}{q}{}^{+i} = \epsilon^{ik}q_{k}^+$. In such a
$\omega$ representation, however, the coupling to $V^{++}$ 
contains explicit harmonics,
which is inconvenient for practical calculations.}. The right-hand
sides in (\ref{12}) do not contain any
dependence  on $V^{++}$ and, hence, the corresponding Jacobian is a constant.
Now, putting (\ref{12}) in path integral (\ref{11}),
and eliminating the auxiliary superfields $f^{++}$ and
$\stackrel{\smile}{f}{}^{++}$, one readily finds
\be
{\tilde\Gamma[V^{++}]}=\Gamma_\omega[V^{++}]\;.
\ee
Comparing this with (\ref{11}) leads to (\ref{9}).
Thus to find the
effective action of the theory (\ref{5}), it is sufficient to calculate the
effective action $\Gamma[V^{++}]$ for the theory (\ref{1})

\vspace{0.5cm}

For the correct definition of the effective action
$\Gamma[V^{++}]$, we
consider the Greens function $G^{(1,1)}(1,2)$ of operator $\nabla^{++}$
\be
\nabla_1^{++}G^{(1,1)}(1,2)=\delta_A^{(3,1)}(1,2)
\label{13}
\ee
where $1,2\equiv (\zeta_{1,2A},u_{1,2})$ and
$\delta_A^{(3,1)}(1,2)$ is the appropriate analytic subspace
$\delta$-function \cite{gios1}. Let us
introduce an analytic superkernel $Q^{(3,1)}(1,2)$ which contains all
information about the interaction and is defined by the rule
\be
G_0^{(1,1)}(1,2)=\int {\rm d}\zeta^{(-4)}_{3A}{\rm d}u_3
G^{(1.1)}(1,3)Q^{(3,1)}(3,2)
\label{14}
\ee
with $G_0^{(1,1)}$ the Greens function of the free hypermultiplet \cite{gios1}
\be
G_0^{(1,1)}(1,2) \equiv <\stackrel{\smile}{q}{}^+(1)q^+(2)> = -{1\over \Box_1}
(D_1^+)^4(D_2^+)^4 \delta^4(x_1-x_2)\delta^8(\theta_1-\theta_2)
{1\over (u^+_1 u^+_2)^3}\;.
\label{qgreen}
\ee
Then we have
\be
Q^{(3,1)}(1,2)=\delta_A^{(3,1)}(1,2)+{\rm i}V^{++}(1)G_0^{(1,1)}(1,2)\;.
\label{15}
\ee

With the use of $Q^{(3,1)}(1,2)$,
effective action $\Gamma[V^{++}]$ can be defined
in the form\footnote{From a formal point
of view, this definition means that
$\Gamma[V^{++}]=-{\rm i}\,{\rm Tr}\;\ln (G^{(1,1)}/G_0^{(1,1)})$
where we have used the fact that the effective action is
always defined up to a constant.}
\be
\Gamma[V^{++}]={\rm i}\,{\rm Tr}\;\ln Q^{(3,1)}\,.
\label{16}
\ee
Here the operation Tr is understood in the sense
\be
{\rm Tr}\, {\cal F}^{q,4-q}=\int {\rm d}\zeta^{(-4)}_{A}{\rm d}u\,
{\cal F}^{(q,4-q)}(1,2)
\label{17}
\ee
for any analytic superkernel ${\cal F}^{(q,4-q)}(1,2)$.
Eqs. (\ref{15}--\ref{17})
show that the effective action (\ref{16}) is well defined within 
perturbation theory.

We can write the effective action $\Gamma[V^{++}]$ as a perturbation series
in powers of the interaction as
\vspace{3mm}

$\Gamma[V^{++}]=\sum\limits_{n=1}^\infty\Gamma_n[V^{++}]={\rm i}^2$\
\wavelin(1,0)\mov(1,0)\vdot\mov(1.35,0){\Circle(0.7)}\mov(2,0)
{$-\,\displaystyle\frac{1}{2}{\rm i}^3$}
\mov(2.7,0){\wavelin(1,0)}\mov(3.7,0)\vdot\mov(4.05,0){\Circle(0.7)}
\mov(4.3,0)\vdot\mov(4.3,0){\wavelin(1,0)}\mov(5.7,0){+}

$+\,\displaystyle\frac{1}{3}{\rm i}^4$\
\wavelin(1,0)\mov(1,0)\vdot\mov(1.35,0){\Circle(0.7)}\mov(1.7,0)
{\wavelin(1,0)}\mov(1.7,0)\vdot\mov(1.35,0.35)\vdot
\mov(1.35,0.35){\wavelin(0,0.7)}\mov(3,0)
{$+\ \dots\ +\,\displaystyle\frac{(-1)^{n+1}}{n}{\rm i}^{n+1}$}
\mov(6.5,0){\Circle(0.7)}\mov(6.2,-0.2)\vdot\mov(6.8,-0.2)\vdot
\mov(6.2,0.2)\vdot\mov(6.8,0.2)\vdot\mov(6.5,-0.35)\vdot
\mov(6.5,-0.35){\wavelin(0,-0.7)}
\mov(6.15,-0.2){\wavelin-(-0.5,-0.5)}
\mov(6.65,-0.2){\wavelin(0.5,-0.5)}
\mov(5.95,0.2){\wavelin-(-0.5,0.5)}
\mov(6.45,0.2){\wavelin(0.5,0.5)}
\mov(5.75,0.5)\vdot\mov(6.05,0.6)\vdot\mov(6.35,0.5)\vdot
\mov(7,0){$+\ \dots$}\mov(8.7,0){(25)}
\vspace{3mm}

\noindent
where the $n$-th term $\Gamma_n[V^{++}]$ is depicted by a supergraph with
$n$ external $V^{++}$ -legs.

Eq. (\ref{16}) leads to the following structure for $\Gamma_n[V^{++}]$
\setcounter{equation}{25}
\be
\Gamma_n[V^{++}]={\rm i}\frac{(-1)^{n+1}}{n}\,{\rm Tr}\;
({\rm i}V^{++}G_0^{(1,1)})^n\;.
\label{19}
\ee
Taking into account the antisymmetry of $G_0^{(1,1)}$ \cite{gios1}, one
observes
that all the coefficients $\Gamma_n$ with odd $n$ are vanishing.
Therefore, only the supergraphs with even numbers of legs
contribute to the effective action.
$\Gamma[V^{++}]$ can be shown to be gauge invariant.
Hence, each coefficient $\Gamma_n$ (\ref{19})
can ,in fact, only depend on the strengths $W$, $\bar W$ in the
low-energy limit.

\vspace{0.5cm}

As was previously pointed out, the theory under consideration
possesses two different phases corresponding to the cases
$V_0^{++} = 0$ and $V_0^{++} \neq 0$.First let us discuss the
$V^{++}_0 = 0$ case.

We begin with a direct calculation of the term $\Gamma_2[V^{++}]$ which,
in the central basis, reads
\begin{eqnarray}
\Gamma_2[V^{++}]&=&-\frac{{\rm i}^3}{2}\int {\rm d}^4x_1 {\rm d}^4\theta^+_1
{\rm d}u_1 {\rm d}^4x_2 {\rm d}^4\theta^+_2 {\rm d}u_2
\frac{1}{\Box_1}(D_1^+)^4(D_2^+)^4
[\delta^4(x_1-x_2)\delta^8(\theta_1-\theta_2)] \nonumber\\
&\times&\frac{1}{\Box_2}(D_2^+)^4(D_1^+)^4
[\delta^4(x_2-x_1)\delta^8(\theta_2-\theta_1)]\frac{V^{++}(x_1,\theta_1,
u_1)V^{++}(x_2,\theta_2,u_2)}{(u_1^+u_2^+)^3(u_2^+u_1^+)^3}
\label{20}
\end{eqnarray}
where the explicit form of $G_0^{(1,1)}$ (\ref{qgreen})
has been used.
\footnote{
We use the following notation:
$(D^{\pm})^2=\frac{1}{4}D^{\pm\alpha}D^{\pm}_\alpha$,
$({\bar D}^{\pm})^2=\frac{1}{4}{\bar D}^{\pm}_{\dot\alpha}
{\bar D}^{\pm \dot{\alpha}}$ and $(D^+)^4=(D^+)^2({\bar D}^+)^2$.}
Let us restore the full Grassmann measure
${\rm d}^8\theta_1{\rm d}^8\theta_2$ \cite{gios2},
make use of the relation between $V^{++}$ and
$V^{--}$ \cite{zup}
$$
V^{--}(x,\theta, u) =
\int {\rm d}u_1 \frac{V^{++}(x, \theta, u_1)}{(u^+u^+_1)^2}\;,
$$
and perform the Fourier transform. As a result one
obtains
\be
\Gamma_2[V^{++}]=-\frac{{\rm i}}{2}\frac{1}{(2\pi)^8}\int {\rm d}^4p
{\rm d}^8\theta
{\rm d}u\,V^{++}(p,\theta,u)V^{--}(-p,\theta,u)\Pi(p)
\label{21}
\ee
where
\be
\Pi(p)=\int\frac{{\rm d}^4q}{q^2(q-p)^2} \;.
\label{22}
\ee
Regularizing $\Gamma[V^{++}]$ by the dimensional regularization
prescription
\be
\Pi(p) \quad \rightarrow \quad
\Pi_{reg}(p)=\mu^{2\varepsilon}\int\frac{{\rm d}^Dq}
{q^2(q^2-p^2)}
\label{23}
\ee
with $D=4-2\varepsilon$ and $\mu$ the normalization parameter, and
subtracting the ultraviolet divergence
\be
\Gamma_{div}[V^{++}]=\frac{1}{32\pi^2\varepsilon}\int {\rm d}^4x
{\rm d}^4\theta W^2
\label{24}
\ee
one ends up with the two-leg correction to the renormalized effective action
$\Gamma_{R}[V^{++}]$
\be
\Gamma_{2\,R}[V^{++}]=-\frac{1}{32\pi^2}\int {\rm d}^4x{\rm d}^4\theta
W\ln\left(
-\frac{\Box}{\mu^2}\right)W \;.
\label{27}
\ee
An analogous quantum correction
has been found in $N=1$ super Yang-Mills theory in
\cite{ky}.
Eq. (\ref{27}) can be treated as the leading term
in the effective action for a weak but rapidly varying gauge superfield.
However, for this correction is problematical in the
low-energy limit
where $p^2\rightarrow 0$. To overcome this, we introduce an infrared cutoff
$\Lambda^2$ using the rule
\be
\Pi_{reg}(0)=\mu^{2\varepsilon}\int\limits
_{\Lambda^2}\frac{{\rm d}^Dq}{q^4}={\rm i}
\pi^2\left(\frac{1}{\varepsilon}+
\ln\frac{\mu^2}{\Lambda^2}\right)\;.
\label{lambda}
\ee
Then, the low-energy correction reads
\be
\Gamma_{2\,R}[V^{++}]=-\frac{1}{32\pi^2}\ln\frac{\Lambda^2}{\mu^2}
\int {\rm d}^4x{\rm d}^4\theta W^2\;.
\label{26}
\ee
Eq. (\ref{24}) constitutes the only
divergence in the theory under consideration. All contributions
$\Gamma_n[V^{++}]$ for $n>2$ are automatically ultraviolet-finite.
Clearly, eq. (\ref{26}) corresponds to a holomorphic contribution to
the effective action.

The next stage is the calculation of the four-leg contribution
$\Gamma_4[V^{++}]$ in the low-energy limit. We start with general relation
(\ref{19}) for $n=4$ and restore the full Grassmann measure
${\rm d}^8\theta$.
As the result, we get
\begin{eqnarray}
&\Gamma_4[V^{++}]=-\displaystyle\frac{{\rm i}}{4}\int {\rm d}^4x_1 {\rm d}^4x_2
{\rm d}^4x_3{\rm d}^4x_4 {\rm d}^8\theta_1{\rm d}^8\theta_2
{\rm d}u_1{\rm d}u_2{\rm d}u_3{\rm d}u_4\displaystyle
\frac{1}{\Box_1}(D^+_1)^4
(D^+_2)^4&\nonumber\\
&\times[\delta^4(x_1-x_2)\delta^8(\theta_1-\theta_2)]
\left(\displaystyle\frac{1}{\Box_2}\delta^4(x_3-x_4)\right)
\displaystyle\frac{1}{\Box_3}
(D^+_3)^4(D^+_4)^4[\delta^4(x_3-x_4)\delta^8(\theta_1-\theta_2)]
& \nonumber\\
&\times\left(\displaystyle\frac{1}{\Box_4}\delta^4(x_4-x_1)\right)
\displaystyle\frac{V^{++}(x_1,\theta_1,u_1)V^{++}(x_2,\theta_2,u_2)
V^{++}(x_3,\theta_3,u_3)V^{++}(x_4,\theta_4,u_4)}
{(u^+_1u^+_2)^3(u^+_2u^+_3)^3(u^+_3u^+_4)^3(u^+_4u^+_1)^3} \;.&
\label{28}
\end{eqnarray}
Here we have used the explicit form of $G_0^{(1,1)}$ (\ref{qgreen}) and
integrated over two Grassmann coordinates.

After performing the Fourier transformation of $\delta$-function, the
previous expression can be
rewritten in the form
\begin{eqnarray}
&\Gamma_4[V^{++}]=-\displaystyle\frac{{\rm i}}{4}
\int {\rm d}^4x_1 {\rm d}^4x_2
{\rm d}^4x_3{\rm d}^4x_4 {\rm d}^8\theta_1{\rm d}^8\theta_2
{\rm d}u_1{\rm d}u_2{\rm d}u_3{\rm d}u_4
\displaystyle\frac{{\rm d}^4p_1 {\rm d}^4p_2 {\rm d}^4p_3 {\rm d}^4p_4}
{(2\pi)^{16} p^2_1 p^2_2 p^2_3 p^2_4}&\nonumber\\
&\times{\rm exp}({\rm i}p_1(x_1-x_2)){\rm exp}({\rm i}p_2(x_2-x_3))
{\rm exp}({\rm i}p_3(x_3-x_4)){\rm exp}({\rm i}p_4(x_4-x_1))&\nonumber\\
&\times\delta^8(\theta_1-\theta_2)V^{++}(x_1,\theta_1,u_1)
V^{++}(x_2,\theta_2,u_2)
[D^{+}_2(-p_1)]^4[D^{+}_1(p_1)]^4&\nonumber\\
&\times\displaystyle\frac{[V^{++}(x_3,\theta_2,u_3)V^{++}(x_4,\theta_1,u_4)
[D^{+}_3(p_3)]^4[D^{+}_4(-p_3)]^4\delta^8(\theta_1-\theta_2)]}
{(u^+_1u^+_2)^3(u^+_2u^+_3)^3(u^+_3u^+_4)^3(u^+_4u^+_1)^3}\;.&
\label{29}
\end{eqnarray}
We have omitted the terms obtained by the action of
$[D^{+}_1(p_1)]^4$ on $V^{++}(x_1,\theta_1,u_1)$
and $[D^{+}_2(-p_1)]^4$ on $V^{++}(x_2,\theta_2,u_2)$ because they
do not contribute in the local limit.

Our aim is to find the local low-energy contribution to $\Gamma_4[V^{++}]$.
Due to the supergauge invariance, it should be
composed only from the superfield strengths $W$ and ${\bar W}$
at the same
point $(x,\theta)$.
This means that we are led to consider $W$, ${\bar W}$ as independent
functional arguments of $\Gamma_4[V^{++}]$, neglecting all space-time
derivatives of these superfields. Taking into account the relation
between $W$ and $V^{++}$, eq. (\ref{104}),
there is only one possible way
to convert all
$V^{++}$ into the superfield strengths. It is necessary to distribute eight
spinor derivatives among the external lines so as to have an equal number of
the derivatives $D^+$ and ${\bar D}^+$ acting on the Grassmann
$\delta$-function; otherwise the result
will be zero. It is evident that we get both $W$ and ${\bar W}$ in this
manner and, hence, a non-holomorphic contribution.

Let us briefly discuss the possibility to obtain holomorphic
contributions. Such a contribution is defined by an integral over the chiral
subspace which can be obtained by the rule
$\int {\rm d}^4x{\rm d}^8\theta\sim\int {\rm d}^4x{\rm d}^4
\theta {\bar D}^4$. Then we could throw only four spinor derivatives
on the external legs and distribute the remainder among the
$\delta$-functions. The total number of these derivatives formally
suffices to obtain a non-zero result. Unfortunately, all derivatives acting
on the external legs should have the same chirality in order to finally get
the expression depending only on $W$.
This means that the numbers of $D^+$'s and ${\bar D}^+$'s acting on
the $\delta$-function do not match each other and the final result
must vanish. Thus, there is no holomorphic contribution to
$\Gamma_4[V^{++}]$.

The only part of $\Gamma_4[V^{++}]$ which contains eight spinor
derivatives on external lines can be singled out as follows
\begin{eqnarray}
&\Gamma_4[V^{++}] \Rightarrow -\displaystyle\frac{{\rm i}}{4}\int {\rm d}^4x_1
{\rm d}^4x_2{\rm d}^4x_3{\rm d}^4x_4 {\rm d}^8\theta_1{\rm d}^8\theta_2
{\rm d}u_1{\rm d}u_2{\rm d}u_3{\rm d}u_4
\displaystyle\frac{{\rm d}^4p_1 {\rm d}^4p_2 {\rm d}^4p_3 {\rm d}^4p_4}
{(2\pi)^{16} p^2_1 p^2_2 p^2_3 p^2_4}&\nonumber\\
&\times{\rm exp}({\rm i}p_1(x_1-x_2)){\rm exp}({\rm i}p_2(x_2-x_3))
{\rm exp}({\rm i}p_3(x_3-x_4)){\rm exp}({\rm i}p_4(x_4-x_1))&\nonumber\\
&\times\delta^8(\theta_1-\theta_2)
[D^{+}_3(p_3)]^4[D^{+}_4(-p_3)]^4\delta^8(\theta_1-\theta_2)
 &\nonumber\\
&\times\displaystyle\frac{V^{++}(x_1,\theta_1,u_1)V^{++}(x_2,\theta_2,u_2)
[D^{+}_2(-p_1)]^4V^{++}(x_3,\theta_2,u_3)
[D^{+}_1(p_1)]^4V^{++}(x_4,\theta_1,u_4)}
{(u^+_1u^+_2)^3(u^+_2u^+_3)^3(u^+_3u^+_4)^3(u^+_4u^+_1)^3}&
\label{30}
\end{eqnarray}

After performing the $D$-algebra and integrating over $\theta_2$
one gets in the low-energy limit
\bea
\Gamma_4[V^{++}]&=&-\frac{{\rm i}}{8(2\pi)^4}\int\limits_{\Lambda^2}
\frac{{\rm d}^4p}{p^8}\int
{\rm d}^4x{\rm d}^8\theta\int {\rm d}u_1(D^-_1)^2V^{++}(x,\theta,u_1)
 \nonumber\\
&\times&\int {\rm d}u_2({\bar D}^-_2)^2V^{++}(x,\theta,u_2)\int {\rm d}u_3
({\bar D}^-_3)^2V^{++}(x,\theta,u_3)\nonumber\\
&\times&\int {\rm d}u_4(D^-_4)^2V^{++}(x,\theta,u_4)
\label{31}
\eea
with $\Lambda^2$ the infrared cutoff. Now let us use the relations
(\ref{104})
which allow us to represent eq. (\ref{31}) in a manifestly gauge invariant
form
\be
\Gamma_4[V^{++}]=\frac{1}{(16\pi)^2\Lambda^4}\int {\rm d}^4x{\rm d}^8\theta
W^2{\bar W}^2 \;.
\label{33}
\ee
This result has a simple physical interpretation.
Let us keep as non-vanishing only the electromagnetic field components
$F_{\mu\nu}$ of $W$ and ${\bar W}$.
Then $\Gamma_4[V^{++}]$ turns into
\be
\Gamma_4[V^{++}]=\frac{1}{(64\pi)^2}\frac{1}{\Lambda^4}\int {\rm d}^4x\{
(F_{\mu\nu}F^{\mu\nu})^2+(F_{\mu\nu}{\tilde F}^{\mu\nu})^2\}
\label{34}
\ee
where ${\tilde F}^{\mu\nu}$ is the dual of $F_{\mu\nu}$. Eq. (\ref{34}) is,
in fact, the first time a non-linear quantum correction to the electromagnetic
Lagrangian has been presented for the $N=2$ theories under consideration. 
This type of correction
was originally discussed by Heisenberg and Euler (see, for
instance, \cite{iz}). Therefore, $\Gamma_4[V^{++}]$ can be interpreted
as the $N=2$ supersymmetric generalization of Heisenberg-Euler Lagrangian.
By construction, $\Gamma_4[V^{++}]$ is given in a manifestly $N=2$
supersymmetric and gauge covariant form.

It is worth noticing (see \cite{wgr,h}) that the functional
$\int {\rm d}^4x{\rm d}^8\theta {\bar W}^2W^2$ rewritten in terms of
$N=1$ superfields contains a contribution with four
spinor derivatives of chiral matter superfields. This kind of one-loop
quantum correction to the effective action has been found in Ref. \cite{bky}
and called the effective potential of auxiliary fields (see also \cite{pw}).

The above consideration can be generalized to give the
$2n$-leg contribution $\Gamma_{2n}[V^{++}]$, for $n=3,4,\ldots$, in
the low-energy approximation
\be
\Gamma_{2n}[V^{++}]\sim \int{\rm d}^4x{\rm d}^8\theta\left(\frac
{{\bar W}W}{2\Lambda^2}\right)^n\ \qquad n>1\;.
\label{79}
\ee
Eqs. (\ref{26}) and (\ref{79}) specify the general form
of low-energy effective
action. We see that the effective action has both holomorphic and
non-holomorphic parts. The holomorphic contribution is simple  and
stipulated by the ultraviolet divergence. The non-holomorphic
contribution has a very special structure; i.e. it depends on $W$ and
${\bar W}$ only via the combination ${\bar W}W$.

To fix the dependence on the arbitrary parameter $\mu$ we should, as usual,
impose some renormalization conditions. The infrared cutoff $\Lambda$,
unlike $\mu$, is a physical parameter which, in accordance with the
status of the Wilsonian effective action \cite{wk}, defines
the physical scale where we study the low energy phenomena.

\vspace{0.5cm}

We now turn to the calculation of the low-energy effective
action for the
case when $V_0^{++} \neq 0$.We start from the four-leg contribution (36).
In order to obtain a holomorphic contribution one should throw two
derivatives $D^{+}$ and two derivatives $\bar D^{+}$ on the external lines.
The only term which gives a contribution in the local limit is
\begin{eqnarray}
&\Gamma_4[V^{++}]\Rightarrow -\displaystyle\frac{{\rm i}}{4}
\int {\rm d}^4x_1
{\rm d}^4x_2{\rm d}^4x_3{\rm d}^4x_4 {\rm d}^8\theta_1{\rm d}^8\theta_2
{\rm d}u_1{\rm d}u_2{\rm d}u_3{\rm d}u_4
\displaystyle\frac{{\rm d}^4p_1 {\rm d}^4p_2 {\rm d}^4p_3 {\rm d}^4p_4}
{(2\pi)^{16} p^2_1 p^2_2 p^2_3 p^2_4}&\nonumber\\
&\times{\rm exp}({\rm i}p_1(x_1-x_2)){\rm exp}({\rm i}p_2(x_2-x_3))
{\rm exp}({\rm i}p_3(x_3-x_4)){\rm exp}({\rm i}p_4(x_4-x_1))&\nonumber\\
&\times\delta^8(\theta_1-\theta_2)
[\bar D^{+}_2(-p_1)]^2[D^{+}_1(p_1)]^2[D^{+}_3(p_3)]^4[D^{+}_4(-p_3)]^4
\delta^8(\theta_1-\theta_2)&\nonumber\\
&\times\displaystyle\frac{V^{++}(x_1,\theta_1,u_1)V^{++}(x_2,\theta_2,u_2)
[D^{+}_2(-p_1)]^2V^{++}(x_3,\theta_2,u_3)
[\bar D^{+}_1(p_1)]^2V^{++}(x_4,\theta_1,u_4)}
{(u^+_1u^+_2)^3(u^+_2u^+_3)^3(u^+_3u^+_4)^3(u^+_4u^+_1)^3}\;.&
\label{40}
\end{eqnarray}

The expression we are interested in can be picked out from (42).
Using the fact that $V^{++}=V_0^{++}+V_1^{++}$ with $V_0^{++}$ given
by eq. (5), we can conclude that the local holomorphic contribution comes
from the following piece of $\Gamma_4[V^{++}]$
\begin{eqnarray}
&\Gamma_4[V^{++}]=-\displaystyle\frac{{\rm i}}{4}\int {\rm d}^4x_1
{\rm d}^4x_2{\rm d}^4x_3{\rm d}^4x_4 {\rm d}^8\theta_1{\rm d}^8\theta_2
{\rm d}u_1{\rm d}u_2{\rm d}u_3{\rm d}u_4
\displaystyle\frac{{\rm d}^4p_1 {\rm d}^4p_2 {\rm d}^4p_3 {\rm d}^4p_4}
{(2\pi)^{16} p^2_1 p^2_2 p^2_3 p^2_4}&\nonumber\\
&\times{\rm exp}({\rm i}p_1(x_1-x_2)){\rm exp}({\rm i}p_2(x_2-x_3))
{\rm exp}({\rm i}p_3(x_3-x_4)){\rm exp}({\rm i}p_4(x_4-x_1))&\nonumber\\
&\times \displaystyle\frac{\delta^8(\theta_1-\theta_2)
[\bar D^{+}_2(-p_1)]^2[D^{+}_1(p_1)]^2[D^{+}_3(p_3)]^4[D^{+}_4(-p_3)]^4
\delta^8(\theta_1-\theta_2)}
{(u^+_1u^+_2)^3(u^+_2u^+_3)^3(u^+_3u^+_4)^3(u^+_4u^+_1)^3}&\nonumber\\
&\times V^{++}(x_1,\theta_1,u_1)V^{++}(x_2,\theta_2,u_2)
[(u^+_2u^+_3)^2 \bar W_0
[\bar D^{+}_1(p_1)]^2V_1^{++}(x_4,\theta_1,u_4)&\nonumber\\
&+(u^+_4u^+_1)^2 W_0 [D^{+}_2(-p_1)]^2V_1^{++}(x_3,\theta_2,u_3)
+(u^+_2u^+_3)^2 \bar W_0 (u^+_4u^+_1)^2 W_0]\;.&
\label{41}
\end{eqnarray}

In the low-energy limit eq. (43) gives rise to the gauge
invariant contribution
\begin{eqnarray}
\Gamma_4[V^{++}]=-\displaystyle\frac{{\rm i}}{4}\int
\displaystyle\frac{{\rm d}^4p}{(2\pi)^4p^6}
\int {\rm d}^4x {\rm d}^4\theta W^3 \bar W_0
 + \displaystyle\frac{{\rm i}}{8}\int
\displaystyle\frac{{\rm d}^4p}{(2\pi)^4p^6}
\int {\rm d}^4x {\rm d}^4\theta W^2 {\bar W}_0 W_0 +
{\rm h.c.}\
\end{eqnarray}
where the identity
\begin{equation}
\int {\rm d}^4x {\rm d}^8\theta {\rm d}u V^{++}V^{--}K(W)=
\int {\rm d}^4x {\rm d}^4\theta W^2 K(W)
\end{equation}
for arbitrary an holomorphic function $K(W)$ has been used.

Analogously, in the $2n$-th order we have
\begin{eqnarray}
\Gamma_{2n}[V^{++}]=-\displaystyle\frac{{\rm i}}{2n}\int
\displaystyle\frac{{\rm d}^4p}{(2\pi)^4p^{2n+2}}
\int {\rm d}^4x {\rm d}^4\theta W^{n+1} {\bar W}_0^{n-1}\nonumber\\
 +\displaystyle\frac{{\rm i}}{4n}\int
\displaystyle\frac{{\rm d}^4p}{(2\pi)^4p^{2n+2}}
\int {\rm d}^4x {\rm d}^4\theta W^2 ({\bar W}_0 W_0)^{n-1}+
{\rm h.c.}\
\end{eqnarray}

To calculate the total one-loop effective action we should sum up
all contributions (46). This leads to the expression
\begin{eqnarray}
\Gamma[V^{++}]=\displaystyle\frac{1}{32\pi^2}\int
{\rm d}^4x {\rm d}^4\theta \frac{W}{\bar W_0}
\int {\rm d}p^2  \ln (1+\frac{W \bar W_0}{p^2})\nonumber\\
 -\displaystyle\frac{1}{64\pi^2}\int
{\rm d}^4x {\rm d}^4\theta \frac{W^2}{W_0 \bar W_0}
\int {\rm d}p^2  \ln (1+\frac{W_0 \bar W_0}{p^2})+{\rm h.c.}
\end{eqnarray}

After renormalization and doing the momentum integral, one gets
\be
\Gamma_R[V^{++}]=\int{\rm d}^4x{\rm d}^4\theta{\cal F}(W)+{\rm h.c.}
\label{145}
\ee
where
\be
{\cal F}(W)=\frac{1}{64\pi^2}W^2\left(1-\ln\frac{W^2}{\mu^2}\;.
\right)
\label{146}
\ee
Here all the dependence on $W_0$, ${\bar W}_0$ has been absorbed in the
normalization point $\mu$.

Eqs. (\ref{145}) and (\ref{146}) are two of the main results of our paper.
We see that
the massive branch of the theory , unlike the massless one, allows one to
obtain non-trivial holomorphic contribution to the low-energy effective
action. This holomorphic contribution does not depend on the infrared
cutoff and, hence, it is automatically infrared-finite. To fix the
ultraviolet normalization point we should impose, as usual, some
renormalization condition such as
\be
{\cal F}(W)|_{W^2=M^2}=0\;.
\label{147}
\ee
It means that the quantum correction to the classical Lagrangian $\frac{1}{2}W^2$
is absent at the scale $M$. The above condition fixes the
normalization point $\mu$ and allows us to rewrite eq. (\ref{146}) in
the form
\be
{\cal F}(W)=-\frac{1}{64\pi^2}W^2\ln\frac{W^2}{M^2}\;.
\label{148}
\ee
It is interesting to note that eq. (\ref{148}) coincides, up to
sign and numerical coefficient, with the perturbative holomorphic quantum
correction to the classical Lagrangian of $N=2$ super Yang-Mills theory
which was found by Seiberg \cite{s}, based on non-manifestly
$N=2$ supersymmetric considerations. The difference in sign
and the coefficient is due to two reasons. Firstly, we compute
the quantum correction coming from matter superfields, not gauge ones,
which leads to the opposite sign of the $\beta$-function. Secondly,
the present model describes different degrees of freedom as compared to the
$N=2$ super Yang-Mills model.

\vspace{0.5cm}

Let us summarize the results. We have developed a general
approach to the problem of computing the effective action of the
$N=2$, $D=4$ abelian
gauge superfield coupled to massless and massive off-shell
hypermultiplets (with the mass arising as an effect of the
non-zero vacuum expectation value of the gauge superfield).
This approach is based on the formulation of $N=2$
supersymmetric theories in harmonic superspace and guarantees manifest
$N=2$ supersymmetry at each step of the computation.  We have demonstrated
that the $N=2$ supergraph techniques of Refs.\cite{gios1,gios2} are suitable
for the investigation of a broad class of $N=2$ supersymmetric
theories in the same way, and with the same degree of efficiency, as the
well known $N=1$ supergraph techniques (see, for instance, \cite{ggrs,bk}).

Theory (1) possesses two different phases corresponding
to massless and massive hypermultiplets.
The renormalized Wilsonian effective action of
the Maxwell multiplet was considered for both phases of the theory.
We calculated its explicit form, which depends 
only on the superfield strengths $W$ and ${\bar
W}$, in the low-energy limit where all
derivatives on the superfield strengths can be neglected. 
In the massless case, we found that the effective action contains the
trivial holomorphic contribution  which is
stipulated by the ultraviolet divergence and the non-trivial
non-holomorphic contributions (\ref{33}) and (\ref{79}). These
non-holomorphic contributions are automatically ultraviolet-finite and
depend on an infrared cutoff $\Lambda$ defining a physical
scale in the theory under consideration. The simplest non-holomorphic
contribution (\ref{33}) leads to the $N=2$ supersymmetric extension of the
well-known Heisenberg-Euler lagrangian.
The massive branch occurs when the hypermultiplet is coupled
to a background gauge superfield $V_0^{++}$ with
the constant strength $W_0 \neq 0$.
$V_0^{++}$ can be associated with the breakdown of the
$U(1)$ factor in the automorphism group $U(1) \times SU(2)$ of $N=2$
supersymmetry.
In the massive case, the structure of the effective action
is changed drastically as compared to the massless case.
Here the effective action contains non-trivial holomorphic
contributions.
Moreover, their structure is analogous to the
low-energy perturbative effective action for $N=2$ super
Yang-Mills theory obtained by Seiberg by integrating
the $R$-anomaly [3].
\vspace{1cm}

\noindent
{\bf Acknowledgements}
I.L.B., E.A.I. and S.M.K. acknowledge a partial support from the
RFBR-DFG project No 96-02-00180. Research of I.L.B. and S.M.K. was
partly supported by RFBR under the project No 96-02-16017. Research
of E.A.I. was partly supported by RFBR under the project No
96-02-17634, by INTAS under the project INTAS-94-2317 and by a grant of
the Dutch NWO organization. I.L.B. is very
grateful to the Department of Physics, University of Pennsylvania where
part of this research was carried out, and
for support from the Research Foundation of the University of Pennsylvania.
B.A.O. acknowledges the DOE Contract No. DE-AC02-76-ER-03072 for partial
support.
We thank B.M. Zupnik for collaboration at the early stage of this work and
valuable discussions.

\end{document}